\documentclass[twoside,twocolumn,9pt]{article}
\usepackage{extsizes}
\usepackage[super,sort&compress,comma]{natbib} 
\usepackage[version=3]{mhchem}
\usepackage[left=1.5cm, right=1.5cm, top=1.785cm, bottom=2.0cm]{geometry}
\usepackage{balance}
\usepackage{mathptmx}
\usepackage{sectsty}
\usepackage{graphicx} 
\usepackage{lastpage}
\usepackage[format=plain,justification=justified,singlelinecheck=false,font={stretch=1.125,small,sf},labelfont=bf,labelsep=space]{caption}
\usepackage{float}
\usepackage{fancyhdr}
\usepackage{fnpos}
\usepackage[english]{babel}
\addto{\captionsenglish}{%
  
}
\usepackage{array}
\usepackage{droidsans}
\usepackage{charter}
\usepackage[T1]{fontenc}
\usepackage[usenames,dvipsnames]{xcolor}
\usepackage{setspace}
\usepackage[compact]{titlesec}
\usepackage{hyperref}
\usepackage{gensymb}

\usepackage{epstopdf}

\definecolor{cream}{RGB}{222,217,201}

\begin{document}

\pagestyle{fancy}
\thispagestyle{plain}
\fancypagestyle{plain}{
\renewcommand{\headrulewidth}{0pt}
}

\makeFNbottom
\makeatletter
\renewcommand\LARGE{\@setfontsize\LARGE{15pt}{17}}
\renewcommand\Large{\@setfontsize\Large{12pt}{14}}
\renewcommand\large{\@setfontsize\large{10pt}{12}}
\renewcommand\footnotesize{\@setfontsize\footnotesize{7pt}{10}}
\makeatother

\renewcommand{\thefootnote}{\fnsymbol{footnote}}
\renewcommand\footnoterule{\vspace*{1pt}%
\color{cream}\hrule width 3.5in height 0.4pt \color{black}\vspace*{5pt}} 
\setcounter{secnumdepth}{5}

\makeatletter 
\renewcommand\@biblabel[1]{#1}            
\renewcommand\@makefntext[1]%
{\noindent\makebox[0pt][r]{\@thefnmark\,}#1}
\makeatother 
\renewcommand{\figurename}{\small{Fig.}~}
\sectionfont{\sffamily\Large}
\subsectionfont{\normalsize}
\subsubsectionfont{\bf}
\setstretch{1.125} 
\setlength{\skip\footins}{0.8cm}
\setlength{\footnotesep}{0.25cm}
\setlength{\jot}{10pt}
\titlespacing*{\section}{0pt}{4pt}{4pt}
\titlespacing*{\subsection}{0pt}{15pt}{1pt}

\fancyfoot{}
\fancyfoot[LO,RE]{\vspace{-7.1pt}\includegraphics[height=9pt]{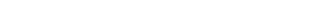}}
\fancyfoot[CO]{\vspace{-7.1pt}\hspace{13.2cm}\includegraphics{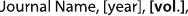}}
\fancyfoot[CE]{\vspace{-7.2pt}\hspace{-14.2cm}\includegraphics{head_foot/RF}}
\fancyfoot[RO]{\footnotesize{\sffamily{1--\pageref{LastPage} ~\textbar  \hspace{2pt}\thepage}}}
\fancyfoot[LE]{\footnotesize{\sffamily{\thepage~\textbar\hspace{3.45cm} 1--\pageref{LastPage}}}}
\fancyhead{}
\renewcommand{\headrulewidth}{0pt} 
\renewcommand{\footrulewidth}{0pt}
\setlength{\arrayrulewidth}{1pt}
\setlength{\columnsep}{6.5mm}
\setlength\bibsep{1pt}

\makeatletter 
\newlength{\figrulesep} 
\setlength{\figrulesep}{0.5\textfloatsep} 

\newcommand{\topfigrule}{\vspace*{-1pt}%
\noindent{\color{cream}\rule[-\figrulesep]{\columnwidth}{1.5pt}} }

\newcommand{\botfigrule}{\vspace*{-2pt}%
\noindent{\color{cream}\rule[\figrulesep]{\columnwidth}{1.5pt}} }

\newcommand{\dblfigrule}{\vspace*{-1pt}%
\noindent{\color{cream}\rule[-\figrulesep]{\textwidth}{1.5pt}} }

\makeatother

\twocolumn[
  \begin{@twocolumnfalse}
{\includegraphics[height=30pt]{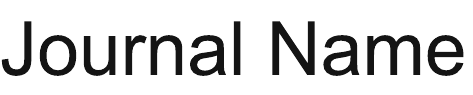}\hfill\raisebox{0pt}[0pt][0pt]{\includegraphics[height=55pt]{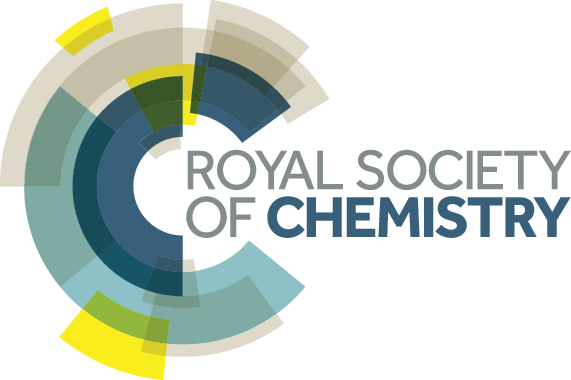}}\\[1ex]
\includegraphics[width=18.5cm]{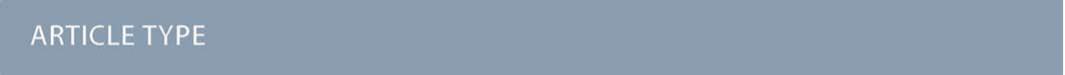}}\par
\vspace{1em}
\sffamily
\begin{tabular}{m{4.5cm} p{13.5cm} }

\includegraphics{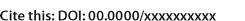} & \noindent\LARGE{\textbf{Composition-Dependent Thermoelectric Properties of Hybrid Tin Perovskites (CH$_3$NH$_3$)$_x$Cs$_{1-x}$SnI$_3$: Insights into Electrical and Thermal Transport Performance} $^\dag $} \\
\vspace{0.3cm} & \vspace{0.3cm} \\

 & \noindent\large{Alexandra Ivanova,$^{\ast}$\textit{$^{a}$} Olga Kutsemako,\textit{$^{a}$} Aleksandra Khanina,\textit{$^{a}$}  Pavel Gorbachev,\textit{$^{a}$} Margarita Golikova,\textit{$^{a}$} Irina Shamova,\textit{$^{a}$} Olga Volkova,\textit{$^{b}$} Lev Luchnikov,\textit{$^{a}$}  Pavel Gostishchev,\textit{$^{a}$} Danila Saranin\textit{$^{a}$} and Vladimir Khovaylo\textit{$^{a}$}} \\

\includegraphics{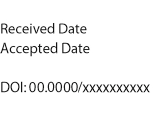} & \noindent\normalsize{This work presents a comprehensive investigation of the thermoelectric properties of bulk hybrid perovskites with the general formula $\text{MA}_x\text{Cs}_{1-x}\text{SnI}_3$ ($0 \leq x \leq 1$). A series of bulk samples were synthesized and systematically characterized to explore the relationship between composition, microstructure, and thermoelectric performance. Compositions with intermediate $\text{MA}^+$ content ($x = 0.2$ and $x = 0.5$) show an optimal balance between electrical conductivity and Seebeck coefficient, yielding high power factor values (0.6--0.7 $\mu$W/cm$\cdot$K$^2$ at 423 K) and favorable thermoelectric performance with $zT$ values up to 0.06. In contrast, compositions with $\text{MA}^+$ contents ($x = 0$, $x = 0.6$, and $x = 0.8$) exhibit lower thermoelectric performance due to reduced Seebeck coefficients or suppressed conductivity. $\text{MASnI}_3$ shows promising low-temperature thermoelectric performance with a maximum $zT$ of 0.10 at 423 K, attributed to its rapidly increasing Seebeck coefficient. These findings highlight the importance of microstructural control and composition optimization in the development of hybrid perovskites for thermoelectric applications.} \\

\end{tabular}

 \end{@twocolumnfalse} \vspace{0.6cm}
]

\renewcommand*\rmdefault{bch}\normalfont\upshape
\rmfamily
\section*{}
\vspace{-1cm}


\footnotetext{\textit{$^{a}$~National University of Science and Technology MISIS (NUST MISIS), Leninsky Av. 4, Moscow, 119049, Russia.}

\textit{$^{b}$~Department of Low Temperature Physics and Superconductivity, Moscow State University, Moscow 119991, Russia.}

E-mail: aivanova@misis.ru}

\footnotetext{\dag~Electronic Supplementary Information (ESI) available: XRD data, Raman data, SEM images and EDX mapping, Electrical transport properties, Thermal transport properties. See DOI: 00.0000/00000000.}


\section*{Introduction}

Thermoelectric materials, capable of converting thermal energy into electrical energy and vice versa, are of significant interest for modern technologies.\cite{wei2020review, han2022room, feng2024low, laghzal2024comprehensive} The efficiency of such materials is quantified by the dimensionless figure of merit, $zT$, defined as:
$zT$={$\alpha^2\sigma{T}$}/{($\kappa_{lat}$ - $\kappa_{el}$)}, 
where $\alpha$ represents the Seebeck coefficient, $\sigma$ denotes electrical conductivity, an $\kappa_{lat}$ and $\kappa_{el}$ are lattice and electronic thermal conductivity, respectively.\cite{ioffe1957} Maximizing 
\textit{zT} is crucial for improving the performance of applications such as thermoelectric cooling, and energy harvesting devices.\cite{han2024advancements} Halide perovskites (PVKs) have become one of the most studied classes of materials over the past decade due to their unique electronic, optical, and thermoelectric properties.\cite{zhou2022recent, liu2023high, lopez2024lead} Among them, tin-based (Sn) hybrid perovskites have attracted particular attention due to their low thermal conductivity, high charge carrier mobility, and the ability to fine-tune electronic properties through chemical composition.\cite{zhou2022recent} However, a limiting factor is the tendency of Sn perovskites to undergo oxidation, where Sn$^{2+}$ is converted to Sn$^{4+}$, leading to defect formation and degradation of thermoelectric performance.\cite{baranwal2022recent}

To address the intrinsic instability and oxidation susceptibility of Sn$^{2+}$ perovskites, researchers have explored mixed cation compositions, including organo-inorganic hybrids such as MA$_x$Cs$_{1-x}$SnI$_3$ (MA = CH$_3$NH$_3$$^+$).\cite{noel2014lead, wang2016organic} Although early studies suggested that MA$^+$ might enhance the stability of these materials, particularly due to its flexibility, which helps accommodate lattice strain and potentially slow the oxidation of Sn$^{2+}$, subsequent research has shown that this effect is limited under operational conditions.\cite{aftab2021review, deendyal2024chronological} However, these limitations highlight the need for a deeper understanding of how mixed cation strategies influence the intrinsic stability and thermoelectric performance of Sn-based perovskites, particularly in bulk form.

To date, research on Sn-based perovskites has been focused predominantly on thin films and single crystals, while bulk materials have been largely overlooked.\cite{byranvand2022tin, baranwal2022recent, liu2023organic, ivanova2024halide} Thin films and single crystals are often more convenient for studying fundamental properties like charge carrier mobility and thermal conductivity, which are crucial for optimizing thermoelectric materials. However, bulk materials are relevant due to their larger volume and potential for scalable production. In this work, our objective is to investigate the bulk properties of the MA$_x$Cs$_{1-x}$SnI$_3$ series, focusing on how the mixed cation composition influences their thermoelectric performance, structural stability, and long-term durability.

\section*{Materials and methods} 

\subsection*{Synthesis of MA$_x$Cs$_{1-x}$SnI$_3$}
The MA$_x$Cs$_{1-x}$SnI$_3$ compounds were synthesized using a mechanochemical ball-milling approach. Stoichiometric mixtures of the starting reagents (MAI granules (with a purity of 99.998 \%, from GratcellSolar), CsI granules (with a purity of 99.998 \%, from LLC Lanhit, Russia) and SnI$_2$ granules (with a purity of 99.999 \%, from LLC Lanhit, Russia)) were loaded into a ball mill and processed at 400 rpm with a ball-to-powder ratio of 1:5. The milling was performed in 20 cycles, each consisting of 5 minutes of active milling followed by a 2-minute pause to prevent overheating. After milling, the resulting powders were cold pressed under a uniaxial stress of 250 MPa for 5 minutes in a cylindrical high-strength stainless steel die with an internal diameter of 10 mm. Then, the compacted samples were sealed in an quartz tubes and pressureless sintered at 433 K for 5 hours under vacuum.

\subsection*{Characterization}

The phase purity and crystal structure of the synthesized MA$_x$Cs$_{1-x}$SnI$_3$ samples were analyzed using X-ray diffraction (XRD) on a TDM-20 diffractometer (Dandong Tongda Science \& Technology Co., Ltd., China) with Cu-K$\alpha$ radiation ($\lambda = 1.5419$ \AA). The morphology and chemical composition of the bulk specimens were examined using scanning electron microscopy (SEM; Vega 3 SB, Tescan, Czech Republic) coupled with energy dispersive X-ray spectroscopy (EDX; x-act, Oxford Instruments, UK).

\subsection*{Transport Property Measurements}
The electrical transport properties, including electrical conductivity ($\sigma$) and the Seebeck coefficient ($\alpha$), were measured using a four-probe method and a differential technique, respectively. Sample preparation for these measurements was carried out in an argon-filled glovebox to prevent oxidation. Consolidated pellets were cut into rectangular bars (3 × 10 × 1.5 mm$^3$) using a hand-held string saw, and the contact surfaces were polished to ensure good electrical contact. The measurements were carried out under a helium atmosphere in the temperature range of 280 to 420 K. The thermal conductivity ($\kappa$) was calculated from the measured thermal diffusivity ($\chi$), specific heat capacity ($C_p$), and density ($d$) using the relationship $\kappa = \chi C_p d$. The density of the samples was determined by the Archimedes principle using isopropyl alcohol as the immersion medium. Thermal diffusivity was measured using the laser flash method on a LFA 447 NanoFlash (Netzsch, Germany). The heat capacity of the mixed MA$_x$Cs$_{1-x}$SnI$_3$ samples was estimated using a linear interpolation between the heat capacity of CsSnI$_3$, calculated using the Debye model, and the experimentally measured heat capacity of MASnI$_3$, obtained via a Physical Property Measurement System (“Quantum Design” Physical Properties Measurements System PPMS-9T). The interpolation assumes ideal mixing behavior, where the heat capacity of the mixed system $C_p(x)$ is given by:

\[
C_p(x) = (1 - x) \cdot C_p(\text{CsSnI}_3) + x \cdot C_p(\text{MASnI}_3),
\]

where $x$ represents the molar fraction of MASnI$_3$ in the mixture. This approach is valid under the assumption that the system exhibits no significant deviations from ideality, such as phase transitions or strong interactions between the components, which could lead to non-linear behavior in the heat capacity. The combined uncertainty for all measurements involved in the calculation of the figure of merit $zT$ was estimated to be 16 \%, taking into account the errors associated with electrical conductivity, Seebeck coefficient, thermal diffusivity, and heat capacity measurements.

\section*{Results and discussion}

\begin{figure}[!t]
\centering
    \includegraphics{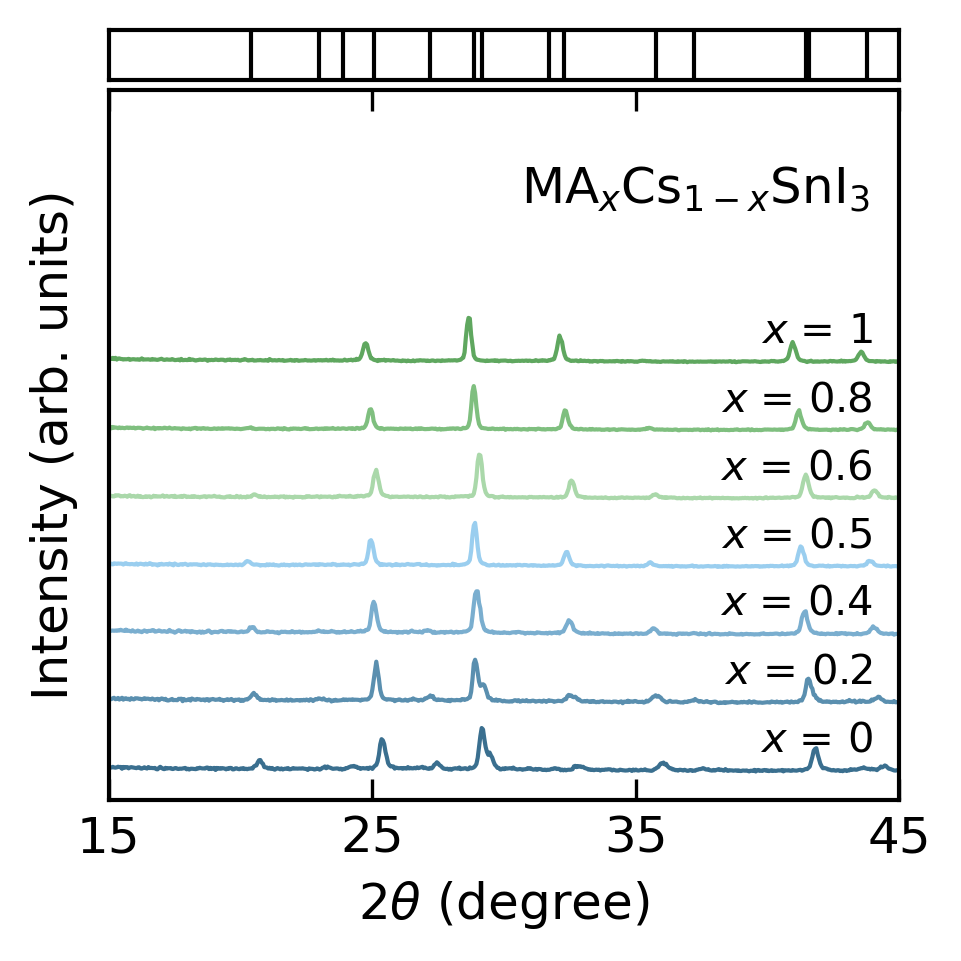}
\caption{X-ray diffraction (XRD) patterns of MA$_x$Cs$_{1-x}$SnI$_3$ for varying compositions (x=0, 0.2, 0.4, 0.5, 0.6, 0.8). The dashed line marks a reference angle for phase comparison.~}
    \label{Fig:XRD}
\end{figure}

X-ray diffraction (XRD) patterns of ball-milled MA$_x$Cs$_{1-x}$SnI$_3$ powder samples ($x = 0$, 0.2, 0.4, 0.5, 0.6, 0.8) confirm the formation of a single-phase orthorhombic perovskite structure (PDF\# 01-080-2139) across the entire composition range (Fig.~S1). A systematic shift of the diffraction peaks toward lower 2$\theta$ angles (e.g., the main peak at $\sim$27\degree) is observed with increasing MA$^+$ content, indicating expansion of the unit cell. This behavior is attributed to the partial substitution of the smaller Cs$^+$ ions (ionic radius $\sim$1.67,\AA) with larger MA$^+$ cations (effective ionic radius $\sim$2.17,\AA), leading to lattice expansion. No secondary phases are detected in any of the samples, confirming the formation of a continuous solid solution and the absence of phase segregation.

A similar trend is observed after pressureless sintering: all MA$_x$Cs$_{1-x}$SnI$_3$ compositions retain the orthorhombic perovskite structure, with diffraction peaks systematically shifting to lower angles (Fig. \ref{Fig:XRD}). Notably, no peaks corresponding to unreacted precursors or possible decomposition products are detected in the sintered samples, indicating complete reaction and high phase purity.

XRD analysis of the samples synthesized by ball milling followed by pressureless sintering (BM+PLS) reveals characteristic degradation behavior (Fig.S2). In the undoped composition ($x = 0$), a rapid transformation is observed from the yellow perovskite phase Y-CsSnI$_3$ (PDF\# 01-071-1898) to the double perovskite Cs$_2$SnI$_6$ (PDF\# 01-073-0330) is observed. The yellow phase remains detectable after 1 hour of ambient exposure, but fully converts to Cs$_2$SnI$_6$ within 6 hours. Compared with the bulk samples synthesized by alternative methods, significant differences in degradation kinetics are observed. The melting followed by spark plasma sintering (VM+SPS) preserves the yellow phase for up to 120 hours, while those prepared by vacuum melting and pressureless sintering (VM+PLS) remain stable for at least 24 hours.\cite{ivanova2024phase, ivanova2025stabilization} These discrepancies are likely related to (i) the lower density of BM+PLS samples, which may accelerate diffusion of oxygen and moisture; (ii) the reduced specific surface area caused by extended ball milling; and (iii) variations in defect concentration and grain boundary characteristics induced by the synthesis route. The stability of VM+SPS samples is attributed to their dense microstructure and low surface area, which restrict the diffusion of oxygen and moisture into the bulk. Moreover, the high-temperature processing involved SPS aids in the thermodynamic stabilization of the yellow phase. In contrast, the nanostructured BM+PLS samples possess a highly developed surface area that facilitates rapid environmental interactions and accelerates the transformation to the double perovskite phase. Additional contributions arise from defect chemistry differences, particularly variations in Sn vacancy concentration and grain boundary density.

A composition-dependent trend is observed in the degradation behavior: samples with MA content above $x > 0.4$ bypass the yellow perovskite phase entirely, undergoing a direct transformation from the orthorhombic to the double perovskite structure. This transition is driven by a combination of stereochemical and thermodynamic effects introduced by large MA$^+$ cations, which induces significant distortions in the SnI$_6$ octahedra, lowers the phase transition energy barrier, and increases hygroscopicity. This factor collectively promotes rapid reorganization into the more stable double perovskite under ambient conditions.\cite{wang2016organic} Although all MA-containing samples predominantly retain the orthorhombic phase after 6 hours of exposure, only MA$_{0.8}$Cs$_{0.2}$SnI$_3$ and MASnI$_3$ remain fully stable during this period. Notably, MA$_{0.8}$Cs$_{0.2}$SnI$_3$ shows nearly complete disappearance of the orthorhombic phase after 24 hours, with only the double perovskite detectable after 48 hours. In contrast, MASnI$_3$ continues to exhibit coexistence of both phases after 24 hours and retains the orthorhombic phase even after 48 hours.

A study of the fracture morphology of Cs$_{1-x}$MA$_x$SnI$_3$ perovskite samples using scanning electron microscopy revealed a pronounced dependence of the microstructure on composition (Fig. S3). For compositions with $x = 0.2$, $0.5$, and $1$, the formation of well-defined crystalline grains with sizes ranging from 1 to 20 $\mu$m and an angular morphology is observed (Fig. S3b, d, g). Notably, XRD analysis of all samples demonstrates identical diffraction patterns despite these morphological differences (Fig. \ref{Fig:XRD}). This indicates that the basic perovskite structure is preserved across the entire composition range, with the morphological variations reflecting surface microstructure and inhomogeneities not detectable by XRD. Compositions with intermediate MA$^+$ content ($x = 0.4$ and $0.8$) exhibit a substantially different morphological pattern (Fig. S3c, f). In these samples, extensive regions with disordered structure dominate, where clear grain boundaries are absent, and the fracture surface appears to be "continuous." EDX analysis confirms the chemical homogeneity of regions in all \ce{MA_x_{1-x}SnI3} compositions ($0 \leq x \leq 1$), ruling out phase impurities (Fig.S4). While elemental mapping was performed for every sample, we showcase representative data for $x = 0$, 0.5, and 1, as all compositions exhibit identical uniformity. The formation of such disordered regions may be linked to kinetic factors during crystallization at these compositions. Of particular interest is the morphology of pure CsSnI$_3$ ($x = 0$) and the composition with $x = 0.6$ (Fig. S3a, e), which exhibit a mixed structure. In addition to monocrystalline blocks characterized by minimal boundaries, these samples contain regions with disordered structure. Notably, the observed morphological differences correlate well with electrical conductivity data, where compositions with more pronounced crystallinity exhibit higher conductivity values compared to samples with extensive disordered regions.

\begin{figure}[!t]
\centering
\includegraphics{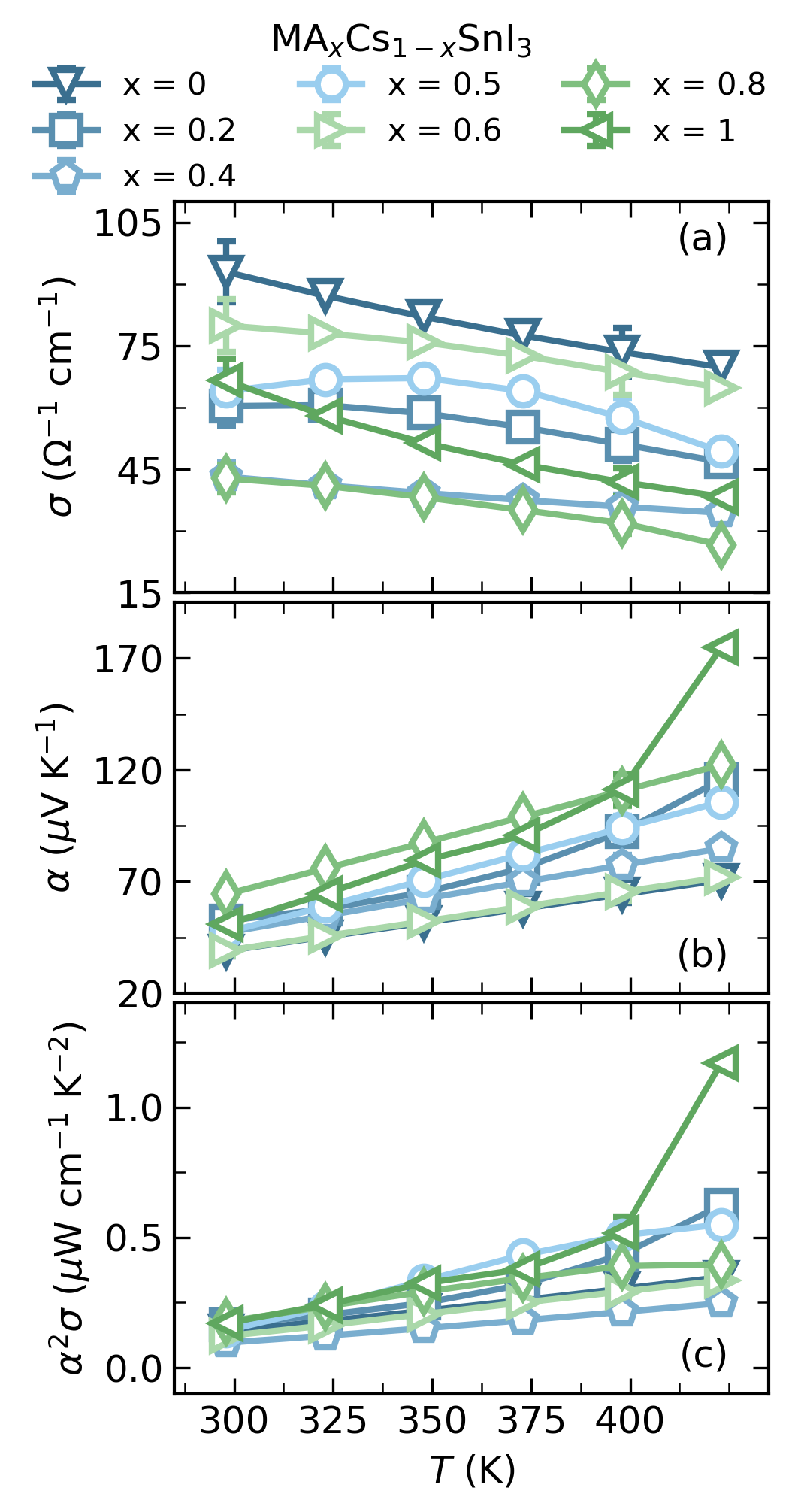}
\caption{Temperature dependence of (a)~the electrical conductivity $\sigma$, 
(b)~the Seebeck coefficient $\alpha$, and 
(c)~the power factor $\alpha^{2}\sigma$ for the 
\ce{MA_xCs_{1-x}SnI3} ($0 \leq x \leq 1$) samples after BM+PLS.}
\label{Fig:eltr}
\end{figure}

The electrical properties of MA$_x$Cs$_{1-x}$SnI$_3$ perovskites exhibit a clear inverse relationship between electrical conductivity and Seebeck coefficient, typical of semiconductor behavior (Fig. \ref{Fig:eltr}). The highest electrical conductivity is observed for $x = 0$ and $x = 0.6$ compositions, with values of 93 and 80~$\Omega^{-1}\mathrm{cm}^{-1}$ at room temperature, respectively (Fig.3a). These compositions also display the lowest Seebeck coefficients of 39~$\mu$V/K (Fig.\ref{Fig:eltr}b), attributed to their well-developed crystalline microstructure (Fig.S3). This microstructure, with large crystalline and minimal grain boundaries, facilitates efficient charge transport but limits thermoelectric voltage generation. Intermediate compositions, such as $x = 0.2$, 0.4, and 0.5, exhibit a broader microstructural mix, containing both crystalline grains and disordered regions. These compositions show conductivity values ranging from 67 to 43~$\Omega^{-1}\mathrm{cm}^{-1}$ and correspondingly higher Seebeck coefficients (47–52~$\mu$V/K), reflecting moderate carrier scattering and enhanced energy-dependent transport phenomena. The $x = 0.8$ composition, with the lowest conductivity (43~$\Omega^{-1}\mathrm{cm}^{-1}$) and the highest Seebeck coefficient (64~$\mu$V/K), demonstrates the most pronounced effect of disordered structure. The absence of long-range crystallinity leads to significant carrier scattering, which suppresses conductivity but enhances thermopower. MASnI$_3$ exhibits non-linear behavior with a sharp increase in Seebeck coefficient from 51~$\mu$V/K at 300~K to 175~$\mu$V/K at 423~K, suggesting the involvement of alternative transport mechanisms, such as thermal activation or an electronic phase transition.

\ce{MASnI3} demonstrates an electrical conductivity that lies within the mid-range of reported literature values (Fig.~S5a). Although some studies report exceptionally high conductivities for bulk materials (200--300~$\Omega^{-1}\mathrm{cm}^{-1}$) \cite{chung2012cssni3, xie2020all}, most findings fall within the 40--120~{$\Omega^{-1}\mathrm{cm}^{-1}$} range \cite{mettan2015tuning, qian2020enhanced, yu2022enhanced, sebastia2022vacuum, tang2022high, wang2022structural, saini2022use, haque2023electrical, tounesi2023influence, ivanova2024halide, ivanova2025stabilization}. At room temperature, \ce{MASnI3} exhibits a relatively low Seebeck coefficient; however, its exponential increase with temperature leads to values that are among the highest reported to date (Fig.~S5c).

The power factor values reach their highest for $x = 0.2$ and $x = 0.5$, with values of 0.6–0.7~$\mu$W cm$^{-1}$K$^{-2}$ at 423~K. These compositions strike an optimal balance between electrical conductivity (60–67~$\Omega^{-1}$cm$^{-1}$) and Seebeck coefficient (47–52~$\mu$V/K). In contrast, the $x = 0$ and $x = 0.6$ compositions, despite exhibiting higher electrical conductivities (93 and 80~$\Omega^{-1}$cm$^{-1}$), show lower power factors (below 0.5~$\mu$W cm$^{-1}$K$^{-2}$) due to their lower Seebeck coefficients. The $x = 0.8$ composition, with the highest Seebeck coefficient (64~$\mu$V/K), exhibits a suppressed power factor owing to its low conductivity (43~$\Omega^{-1}$cm$^{-1}$). \textbf{Additionally, MASnI$_3$ ($x = 1$) shows a significantly lower power factor of 0.1~$\mu$W cm$^{-1}$K$^{-2}$, despite its high Seebeck coefficient.}

Fig. S4a shows the experimental heat capacity data for \ce{MASnI3} and \ce{MAPbI3}, along with the calculated Debye model values for \ce{CsSnI3} in the temperature range of 260--350~K. The measurements for \ce{MASnI3}, obtained using a PPMS system, show good agreement with literature data for \ce{MAPbI3} reported by Ge et al.\cite{ge2018ultralow}. The \ce{MASnI3} curve exhibits an anomaly at a lower temperature (275 K) compared to \ce{MAPbI3} (332 K), indicating an earlier phase transition in the Sn-based perovskite. The calculated heat capacity for \ce{CsSnI3}, derived from the Debye model, is also presented for comparison. Fig.~S4b presents interpolated heat capacity curves for the entire \ce{MA_xCs_{1-x}SnI3} series ($x = 0$--1). The interpolation, based on experimental data from the two end-member compositions (\ce{CsSnI3} and \ce{MASnI3}), captures their respective temperature dependencies. As the \ce{MA^+} content increases, the density of the samples decreases from 4.46 g/cm$^3$ to 3.46 g/cm$^3$, which can be attributed to the substitution of denser \ce{Cs^+} ions with larger, less dense \ce{MA^+} ions, as well as the increase in porosity and dynamic disorder associated with the organic cations.

\begin{figure}[!t]
\centering
    \includegraphics{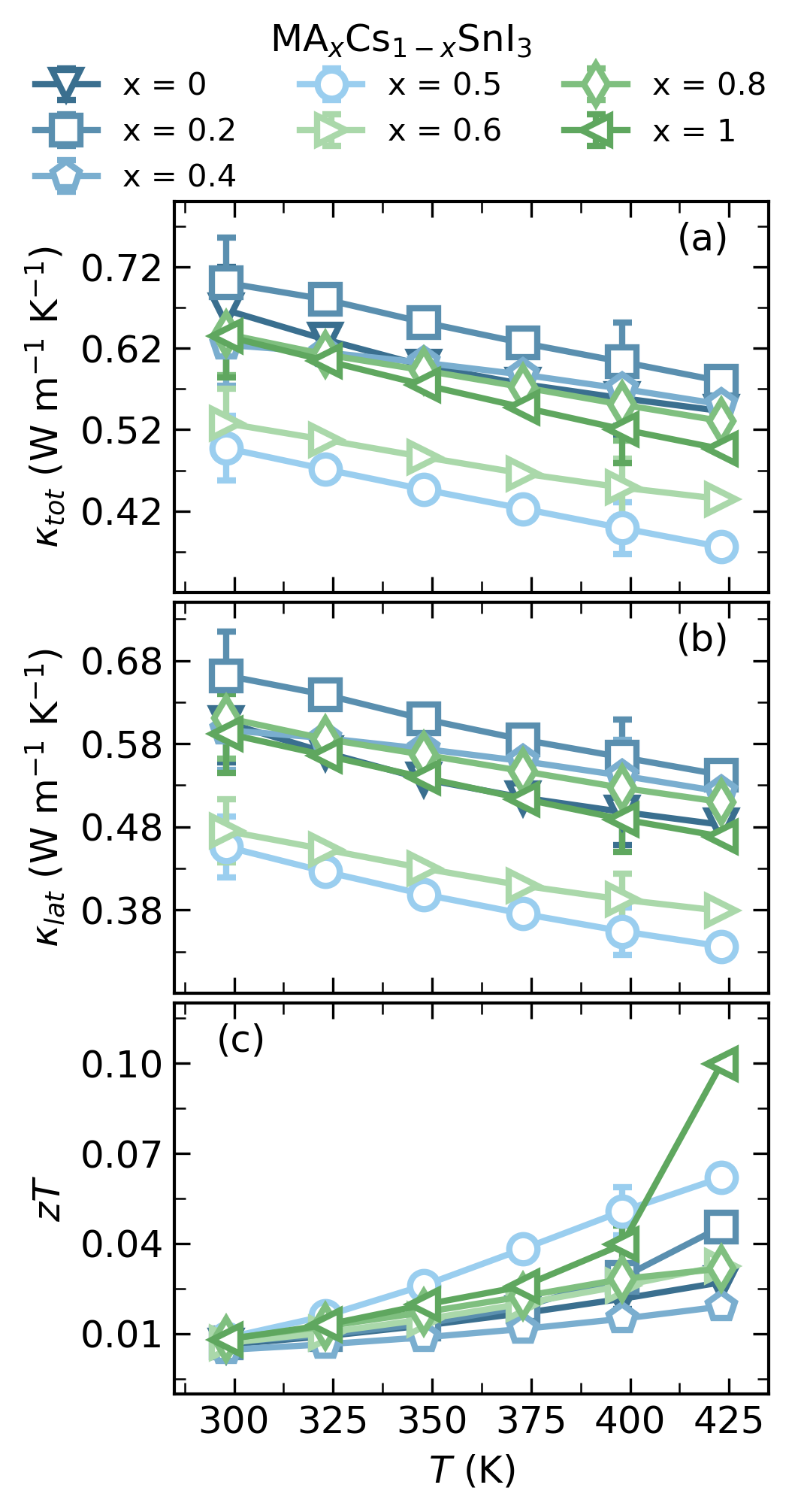}
\caption{Temperature dependence of (a) the total thermal conductivity \(\kappa_{tot}\), (b) the lattice thermal conductivity \(\kappa_{lat}\) and (c) the thermoelectric efficiency \textit{zT} for the \ce{MA_xCs_{1-x}SnI3} ($0 \leq x \leq 1$) samples after BM+PLS.~}
    \label{Fig:zt}
\end{figure}

The lowest thermal conductivity values (Fig.\ref{Fig:zt}a) were observed for the compositions with $x = 0.5$ and $x = 0.6$, measuring 0.50 and 0.53 W m$^{-1}$K$^{-1}$ at room temperature, respectively. SEM analysis of these samples (Fig.~S3) reveals increased porosity and heterogeneous microstructures, which likely contribute to phonon scattering and reduction of thermal conductivity. The presence of pores and grain boundaries can serve as effective scattering centers for phonons, thereby reducing the lattice thermal conductivity component ($\kappa_\mathrm{lat}$).\cite{smith2013thermal} The partial substitution of Cs$^+$ with larger and dynamically disordered MA$^+$ cations may contribute to the reduction of $\kappa_\mathrm{lat}$ through enhanced phonon scattering. In lead-based hybrid perovskites, such as CH$_3$NH$_3$PbI$_3$, the presence of the organic MA$^+$ cation has been shown to induce resonant phonon scattering associated with its rotational motion, effectively suppressing thermal conductivity compared to the fully inorganic counterpart CsPbI$3$.\cite{kovalsky2017thermal} Although a detailed analysis of phonon dynamics in tin-based perovskites is still lacking, a similar mechanism may be operative in the MA$_x$Cs${1-x}$SnI$_3$ system, where mass fluctuation, local lattice strain, and possible dynamic disorder introduced by MA$^+$ could contribute to phonon scattering and lattice thermal conductivity reduction.

Slightly higher thermal conductivities of approximately 0.63 W m$^{-1}$K$^{-1}$ were observed for the compositions with $x = 0.4$, 0.8, and 1.0. In these samples, SEM images show more compact microstructures compared to $x = 0.5$ and 0.6, suggesting reduced phonon scattering at structural defects and improved phonon transport. The composition with $x = 0$ (CsSnI$_3$) exhibits thermal conductivity of 0.67 W m$^{-1}$K$^{-1}$, consistent with its relatively dense microstructure. The highest thermal conductivity in the series, 0.70~W/m·K, was recorded for $x = 0.2$. SEM analysis shows a well-sintered microstructure with minimal porosity and large, interconnected grains (Fig.~S3b), which may facilitate phonon transport by reducing boundary and pore scattering. Prior to this study, thermal conductivity data for bulk MASnI$_3$ had not been reported. Moreover, the proportion of studies addressing the thermal transport properties of perovskites remains critically low. Based on comparison with available literature, the thermal conductivity values obtained for MASnI$_3$ in this work show good agreement with those reported for CsSnI$_3$ (Fig. 5b), suggesting similar phonon transport behavior in these structurally related materials. 

The temperature-dependent thermoelectric figure of merit (\textit{zT}) for all MA$_x$Cs$_{1-x}$SnI$_3$ compositions is shown in Fig.\ref{Fig:zt}c. Samples with MA content of $x = 0$, 0.4, 0.6, and 0.8 exhibit relatively low \textit{zT} values, staying below 0.04 across the temperature range due to the combination of moderate Seebeck coefficients and suboptimal electrical conductivity or thermal transport. The $x = 0.5$ composition, with the lowest thermal conductivity in the series, achieves a higher \textit{zT} of 0.06. The best thermoelectric performance is observed for MASnI$_3$, where a rapidly increasing Seebeck coefficient at elevated temperatures compensates for its moderate electrical and thermal conductivity, yielding a maximum \textit{zT} of 0.10 at 423 K. This value represents one of the highest thermoelectric efficiencies reported for pristine perovskites (Fig. S5d), suggesting that further compositional engineering may lead to even greater performance improvements. 

\section*{Conclusions}

In this study, we systematically investigated the thermoelectric properties of bulk MA$_x$Cs$_{1-x}$SnI$_3$ perovskites, marking a significant step forward in understanding the thermoelectric performance of these materials in their bulk form, as opposed to the more commonly studied thin films and single crystals. Compositions with intermediate MA$^+$ content (\textit{x} = 0.2 and \textit{x} = 0.5) exhibit an optimal balance between electrical conductivity and Seebeck coefficient, resulting in the highest power factor values of 0.6–0.7~$\mu$W cm$^{-1}$K$^{-2}$ at 423~K. These compositions also achieve favorable thermoelectric figure of merit (\textit{zT}) values, with \textit{x} = 0.5 reaching 0.06, indicating their potential for thermoelectric energy conversion in practical bulk applications. 

In contrast, compositions with $x = 0$, $x = 0.6$, and $x = 0.8$ exhibit either low Seebeck coefficients or significantly suppressed conductivity, limiting their thermoelectric performance. Notably, MASnI$_3$, despite its poor performance at lower temperatures, demonstrates promising low-temperature behavior with a rapidly increasing Seebeck coefficient and a \textit{zT} value reaching 0.10 at 423 K, positioning it as a potential candidate for bulk thermoelectric applications. 

\section*{Author Contributions}
\textbf{Alexandra Ivanova}: Conceptualization, Methodology, Formal analysis, Investigation, Data curation, Visualization, Writing (Original draft), Writing (Review \& Editing). \textbf{Olga Kutsemako}: Invesigation. \textbf{Aleksandra Khanina}: Investigation, Data curation. \textbf{Pavel Gorbachev}: Investigation. \textbf{Margarita Golikova}: Investigation. \textbf{Irina Shamova}: Investigation, Data curation. \textbf{Olga Volkova}: Investigation. \textbf{Lev Luchnikov}: Methodology. \textbf{Pavel Gostishchev}: Project administration, Funding acquisition. \textbf{Danila Saranin}: Resources, Supervision, Writing (Review \& Editing). \textbf{Vladimir Khovaylo}: Writing (Review \& Editing), Supervision.

\section*{Conflicts of interest}
There are no conflicts to declare.

\section*{Acknowledgements}
The study was carried out with financial support from the Russian Science Foundation (project no. 22-79-10326).



\balance


\bibliography{rsc} 
\bibliographystyle{rsc} 

\end{document}